\documentclass[draft,tightenlines,nofootinbib,preprint,aps,eqsecnum,amsmath,amssymb]{revtex4}

\newcommand{\beq}{\begin{equation}}
\newcommand{\eeq}{\end{equation}}
\newcommand{\bea}{\begin{eqnarray}}
\newcommand{\eea}{\end{eqnarray}}

\begin{document}

\title{The Chrono-Geometrical Structure of General Relativity and Clock Synchronization.}

\medskip

\author{Luca Lusanna}

\affiliation{ Sezione INFN di Firenze\\ Polo Scientifico\\ Via Sansone 1\\
50019 Sesto Fiorentino (FI), Italy\\  E-mail: lusanna@fi.infn.it}

\begin{abstract}

After a review of the chrono-geometrical structure of special
relativity, where the definition of the instantaneous 3-space is
based on the observer-dependent convention for the synchronization
of distant clocks, it is shown that in a class of models of general
relativity the instantaneous 3-space and the associated clock
synchronization convention are dynamically determined by Einstein's
equations. This theoretical framework is necessary to understand the
relativistic effects around the Earth, to be tested with the ACES
mission of ESA, and the implications for metrology induced by the
accuracy of the new generation of atomic clocks.
\bigskip

Talk at the First Colloquium Scientific and Fundamental Aspects of
the Galileo Programme, Toulouse 1-4 October 2007.

\end{abstract}

\today

\maketitle

\vfill\eject

In Newton physics there are distinct absolute notions of {\it time}
and {\it space}, so that we can speak of absolute simultaneity and
of instantaneous Euclidean 3-spaces with the associated Euclidean
spatial distance notion. This non-dynamical chrono-geometrical
structure is formalized in the so called Galilei space-time. The
Galilei relativity principle assumes the existence of preferred
inertial frames with inertial Cartesian coordinates centered on
inertial observers, connected by the kinematical group of Galilei
transformations. In Newton gravity the equivalence principle states
the equality of inertial and gravitational mass. In non-inertial
frames inertial (or fictitious) forces proportional to the mass of
the body appear in Newton's equations.
\bigskip

In special relativity the structure of the light-cones is an
absolute non-dynamical object \cite{1b}: they are the  only
information (the conformal structure) given by the theory to an
(either inertial or accelerated) observer in each point of her/his
world-line. There is no notion of instantaneous 3-space, of spatial
distance and of one-way velocity of light between two observers .
The light postulates say that the two-way (or round trip) velocity
of light $c$ (only one clock is needed in its definition) is
constant and isotropic. For an ideal inertial observer Einstein's
convention for the synchronization of distant clocks \footnote{The
inertial observer $A$ sends a ray of light at $x^o_i$ to a second
accelerated observer B, who reflects it towards A. The reflected ray
is reabsorbed by the inertial observer at $x^o_f$. The convention
states that the clock of B at the reflection point must be
synchronized with the clock of A when it signs ${1\over 2}\, (x^o_i
+ x^o_f) = x^o_i + {1\over 2}\, (x^o_f - x^o_i)$.} selects the
constant time hyper-planes of the inertial frame having the observer
as time axis as the instantaneous Euclidean 3-spaces, with their
Euclidean 3-geodesic spatial distance and with the one-way and
two-way velocities of light equal.
\medskip

But this convention does not work for realistic accelerated
observers, because coordinate singularities are produced in the
attempt (the 1+3 point of view) to build (either Fermi or rotating)
4-coordinates around the observer world-line. They must use the more
complex conventions arising from the introduction of an extra
structure: a global 3+1 splitting of Minkowski space-time (a choice
of {\it time}, the starting point of the Hamiltonian formalism).
Each space-like leaf of the associated foliation is {\it both a
Cauchy surface for the field equations and a convention} (different
from Einstein's one) {\it for clock synchronization}. If we
introduce {\it Lorentz-scalar observer-dependent radar
4-coordinates} $x^{\mu} \mapsto \sigma^A = (\tau ; \sigma^r)$, where
$x^{\mu}$ are Cartesian coordinates, $\tau$ is an arbitrary
monotonically increasing function of the proper time of the observer
and $\sigma^r$ curvilinear 3-coordinates having the observer
world-line as origin, this leads to the definition of a {\it
non-inertial frame} centered on the accelerated observer \cite{2b}.
Every 3+1 splitting, satisfying certain M$\o$ller restrictions (to
avoid coordinate singularities) and with the leaves tending to
hyper-planes at spatial infinity (so that there are asymptotic
inertial observers to be identified with the fixed stars), gives a
conventional definition of instantaneous 3-space (in general a
Riemannian 3-manifold), of 3-geodesic spatial distance and of
one-way velocity of light (in general both point-dependent and
anisotropic). The inverse coordinate transformation $\sigma^A
\mapsto x^{\mu} = z^{\mu}(\tau ,\sigma^r)$ defines the {\it
embedding} of the simultaneity surfaces $\Sigma_{\tau}$ into
Minkowski space-time. The 3+1 splitting leads to the following
induced 4-metric (a functional of the embedding): ${}^4g_{AB}(\tau ,
\sigma^r) = {{\partial z^{\mu}(\sigma )}\over {\partial \sigma^A}}\,
{}^4\eta_{\mu\nu}\, {{\partial z^{\nu}(\sigma )}\over {\partial
\sigma^B}} = {}^4g_{AB}[z(\sigma )]$.

\medskip

Parametrized Minkowski theories \cite{3b}, \cite{1b} allow to give a
description of every isolated system (particles, strings, fields,
fluids), in which the transition from a 3+1 splitting to another one
(i.e. a change of clock synchronization convention) is a {\it gauge
transformation}. Given any isolated system admitting a Lagrangian
description, one makes the coupling of the system to an external
gravitational field and then replaces the 4-metric
${}^4g_{\mu\nu}(x)$ with the induced metric ${}^4g_{AB}[z(\tau
,\sigma^r)]$ associated to an arbitrary admissible 3+1 splitting.
The Lagrangian now depends not only on the matter configurational
variables but also on the embedding variables $z^{\mu}(\tau
,\sigma^r)$ (whose conjugate canonical momenta are denoted
$\rho_{\mu}(\tau ,\sigma^r)$). Since the action principle turns out
to be invariant under {\it frame-preserving diffeomorphisms}, at the
Hamiltonian level there are four first-class constraints ${\cal
H}_{\mu}(\tau ,\sigma^r) = \rho_{\mu}(\tau ,\sigma^r) - l_{\mu}(\tau
,\sigma^r)\, T^{\tau\tau}(\tau ,\sigma^r) - z^{\mu}_s(\tau
,\sigma^r)\, T^{\tau s}(\tau ,\sigma^r) \approx 0$ in strong
involution with respect to Poisson brackets, $\{ {\cal H}_{\mu}(\tau
,\sigma^r), {\cal H}_{\nu}(\tau ,\sigma_1^r)\} = 0$. Here
$l_{\mu}(\tau ,\sigma^r)$ are the covariant components of the unit
normal to $\Sigma_{\tau}$, while $z^{\mu}_s(\tau ,\sigma^r) =
{{\partial\, z^{\mu}(\tau ,\sigma^r)}\over {\partial\, \sigma^s}}$
are the components of three independent vectors tangent to
$\Sigma_{\tau}$. The quantities $T^{\tau\tau}$ and $T^{\tau s}$ are
the components of the energy-momentum tensor of the matter inside
$\Sigma_{\tau}$ describing its energy- and momentum- densities.
Since the first class constraints are generators of Hamiltonian
gauge transformations, this implies that the configuration variables
$z^{\mu}(\tau ,\sigma^r)$ are arbitrary {\it gauge variables}.
Therefore, all the admissible 3+1 splittings, namely all the
admissible conventions for clock synchronization, and all the
admissible non-inertial frames centered on time-like observers are
{\it gauge equivalent}.\medskip

By adding four gauge-fixing constraints $\chi^{\mu}(\tau ,\sigma^r)
= z^{\mu}(\tau ,\sigma^r) - z^{\mu}_M(\tau ,\sigma^r) \approx 0$
($z^{\mu}_M(\tau ,\sigma^r)$ being an admissible embedding),
satisfying the orbit condition $det\, |\{\chi^{\mu}(\tau ,\sigma^r),
{\cal H}_{\nu}(\tau ,\sigma_1^r)| \not= 0$, we identify the
description of the system in the associated non-inertial frame
centered on a given time-like observer. The resulting effective
Hamiltonian for the $\tau$-evolution turns out to contain the
potentials of the {\it relativistic inertial forces} present in the
given non-inertial frame. As a consequence, the gauge variables
$z^{\mu}(\tau ,\sigma^r)$ describe the {\it spatio-temporal
appearances} of the phenomena in non-inertial frames, which, in
turn, are associated to {\it extended} physical laboratories using a
metrology for their measurements compatible with the notion of
simultaneity (distant clock synchronization convention) of the
non-inertial frame (think to the description of the Earth given by
GPS). Therefore, notwithstanding mathematics tends to use only
coordinate-independent notions, physical metrology forces us to
consider intrinsically coordinate-dependent quantities like the
non-inertial Hamiltonians. For instance, the motion of satellites
around the Earth is governed by a set of empirical coordinates
contained in the software of NASA computers \cite{4b}: this is a
{\it metrological standard of space-time around the Earth}.
\medskip

Inertial frames centered on inertial observers are a special case of
gauge fixing in parametrized Minkowski theories. For each
configuration of an isolated system there is an special 3+1
splitting associated to it: the foliation with space-like
hyper-planes orthogonal to the conserved time-like 4-momentum of the
isolated system. This identifies an intrinsic inertial frame, the
{\it rest-frame}, centered on a suitable inertial observer (the
covariant non-canonical Fokker-Pryce center of inertia of the
isolated system) and allows to define the {\it Wigner-covariant
inertial rest-frame instant form of dynamics} for every isolated
system, which allows to give a new formulation of the relativistic
kinematics \cite{5b} of N-body systems and continuous media
(relativistic centers of mass and canonical relative variables,
rotational kinematics and dynamical body frames, multipolar
expansions, M$\o$ller radius) and to find the theory of relativistic
orbits. Instead {\it non-inertial rest frames} are 3+1 splittings of
Minkowski space-time having the associated simultaneity 3-surfaces
tending to Wigner hyper-planes at spatial infinity.

\bigskip

Instead  in general relativity there is no absolute notion
\cite{1b}: the full chrono-geometrical structure of space-time is
dynamical. The relativistic description of gravity abandons the
relativity principle and replaces it with the equivalence principle.
Special relativity can be recovered only locally by a freely falling
observer in a neighborhood where tidal effects are negligible. As a
consequence, {\it global inertial frames do not exist}.
\medskip

The general covariance of Einstein's formulation of general
relativity leads to a type of gauge symmetry acting also on
space-time: the Hilbert action is invariant under coordinate
transformations ({\it passive} off-shell diffeomorphisms as local
Noether transformations). As in parametrized Minkowski theories the
gauge variables are arbitrary degrees of freedom connected with the
{\it appearances} of phenomena in the various coordinate systems of
Einstein's space-times.

\medskip

In Einstein's geometrical view of the gravitational field the basic
configuration variable is the metric tensor over space-time (10
fields), which, differently from every other field, has a double
role:

i) it is the mediator of the gravitational interaction, like every
other gauge field;

ii) it describes the dynamical chrono-geometrical structure of
space-time by means of the line element $ds^2 = {}^4g_{\mu\nu}(x)\,
dx^{\mu}\, dx^{\nu}$. As a consequence, it {\it teaches relativistic
causality} to the other fields: now the conformal structure (the
allowed  paths of light rays) is point-dependent.

\bigskip

In canonical ADM metric gravity \cite{6b} (and in its tetrad gravity
extension \cite{7b} needed for fermions  \footnote{This leads to an
interpretation of gravity based on a congruence of time-like
observers endowed with orthonormal tetrads: in each point of
space-time the time-like axis is the  unit 4-velocity of the
observer, while the spatial axes are a (gauge) convention for
observer's gyroscopes.}) we have again to start with the same
pattern of 3+1 splittings, to be able to define the Cauchy and
simultaneity surfaces for Einstein's equations. As a consequence,
and having in mind the inclusion of particle physics, we must select
a family of {\it non-compact} space-times $M^4$ with the following
properties:\hfill\break
 i) {\it globally hyperbolic} and {\it topologically trivial},
so that they can be foliated with space-like hyper-surfaces
$\Sigma_{\tau}$ diffeomorphic to $R^3$;\hfill\break
 ii) {\it asymptotically flat at spatial infinity} and with
boundary conditions at spatial infinity independent from the
direction, so that the {\it spi group} of asymptotic symmetries is
reduced to the Poincare' group with the ADM Poincare' charges as
generators. In this way we can eliminate the {\it
super-translations}, namely the obstruction to define angular
momentum in general relativity. All these requirements imply that
the {\it admissible foliations} of space-time must have the
space-like hyper-surfaces tending in a direction-independent way to
Minkowski space-like hyper-planes at spatial infinity, which
moreover must be orthogonal there to the ADM 4-momentum. Therefore,
$M^4$ is {\it asymptotically Minkowskian} \cite{8b} with the
asymptotic Minkowski metric playing the role of an {\it asymptotic
background}. In absence of matter the class of
Christodoulou-Klainermann space-times \cite{9b}, admitting
asymptotic ADM Poincare' charges and an asymptotic flat metric meets
these requirements.

\bigskip

This formulation, the {\it rest-frame instant form of metric and
tetrad gravity}, emphasizes the role of {\it non-inertial frames}
(the only ones existing in general relativity): each admissible 3+1
splitting identifies a global non-inertial frame centered on a
time-like  observer. In these space-times each simultaneity surface
is the rest frame of the 3-universe, there are asymptotic inertial
observers (the fixed stars) and the switching off of the Newton
constant in presence of matter leads to a deparametrization of these
models of general relativity to the non-inertial rest-frame instant
form of the same matter with the ADM Poincare' charges collapsing
into the usual kinematical Poincare' generators. This class of
space-times is suitable to describe the solar system (or the
galaxy), is compatible with particle physics and allows to avoid the
splitting of the metric into a background one plus a perturbation.
With the addition of suitable asymptotic terms it can probably be
adapted to cosmology \cite{10b}.

\medskip

The first-class constraints of canonical gravity (8 in metric
gravity, 14 in tetrad  gravity \footnote{ Tetrad gravity has 10
primary first class constraints and 4 secondary first class ones.
Six of the primary constraints describe the extra freedom in the
choice of the tetrads. The other 4 primary (the vanishing of the
momenta of the lapse and shift functions) and 4 secondary (the
super-Hamiltonian and super-momentum constraints) constraints are
the same as in metric gravity.}) imply the existence of an equal
number of arbitrary gauge variables and of only 2+2 genuine physical
degrees of freedom of the gravitational field: $r_{\bar a}(\tau
,\sigma^r)$, $\pi_{\bar a}(\tau ,\sigma^r)$. It can be shown
\cite{6b,7b,11b} that the super-hamiltonian constraint generates
Hamiltonian gauge transformations implying the {\it gauge
equivalence} of clock synchronization conventions like it happens in
special relativity (no Wheeler-DeWitt interpretation of it as a
Hamiltonian). As shown in Refs.\cite{11b}, the gauge variables
describe {\it generalized inertial effects} (the appearances), while
the 2+2 gauge invariant DO describe {\it generalized tidal effects}.
In Refs.\cite{7b} a  canonical basis, adapted to 13 of the 14 tetrad
gravity first class constraints (not to the super-Hamiltonian one)
was found. With its help it can be shown \cite{11b} that a
completely fixed Hamiltonian gauge is equivalent to the choice of a
{\it non-inertial frame} with its adapted radar coordinates centered
on an accelerated observer and its instantaneous 3-spaces
(simultaneity surfaces): again this corresponds to an extended
physical laboratory \footnote{Let us remark that, if we look at
Minkowski space-time as a special solution of Einstein's equations
with $r_{\bar a}(\tau ,\sigma^r) = \pi_{\bar a}(\tau ,\sigma^r) = 0$
(zero Riemann tensor, no tidal effects, only inertial effects), we
find \cite{6b} that the dynamically admissible 3+1 splittings
(non-inertial frames) must have the simultaneity surfaces
$\Sigma_{\tau}$ {\it 3-conformally flat}, because the conditions
$r_{\bar a}(\tau ,\sigma^r) = \pi_{\bar a}(\tau ,\sigma^r) = 0$
imply the vanishing of the Cotton-York tensor of $\Sigma_{\tau}$.
Instead, in special relativity, considered as an autonomous theory,
all the non-inertial frames compatible with the M$\o$ller conditions
are admissible \cite{5b}, namely there is much more freedom in the
conventions for clock synchronization.}.

\medskip

In the rest-frame instant form of gravity \cite{6b,7b}, due to the
DeWitt surface term the effective Hamiltonian is not weakly zero
({\it no frozen picture} of dynamics), but is given by the weak ADM
energy $E_{ADM} = \int d^3\sigma\, {\cal E}_{ADM}(\tau ,\sigma^r)$
(it is the analogous of the definition of the electric charge as the
volume integral of the charge density in electromagnetism). The ADM
energy density depends on the gauge variables, namely it is a
coordinate-dependent quantity (the {\it problem of energy} in
general relativity). In a completely fixed gauge, in which the
inertial effects are given functions  of the DO, ${\cal
E}_{ADM}(\tau ,\sigma^r)$ becomes a well defined function only of
the DO's and there is a deterministic evolution of the DO's (the
tidal effects) given by the Hamilton  equations. A universe $M^4$ (a
4-geometry) is the equivalence class of all the completely fixed
gauges with gauge equivalent Cauchy data for the DO on the
associated Cauchy and simultaneity surfaces $\Sigma_{\tau}$. In each
completely fixed gauge (an off-shell non-inertial frame determined
by some set of gauge-fixing constraints determining the gauge
variables in terms of the tidal ones) we find the solution for the
DO in that gauge (the tidal effects) and then the explicit form of
the gauge variables (the inertial effects). As a consequence, the
final admissible (on-shell gauge equivalent) non-inertial frames
associated to a 4-geometry (and their instantaneous 3-spaces, i.e.
their clock synchronization conventions) are {\it dynamically
determined} \cite{11b}.

\bigskip

A first application of this formalism  has been the determination
(see the third paper in Refs.\cite{7b}) of {\it post-Minkowskian
background-independent gravitational waves} in a completely fixed
non-harmonic 3-orthogonal gauge with diagonal 3-metric by adding the
weak field requirement $r_{\bar a}(\tau ,\sigma^r) << 1$, $\pi_{\bar
a}(\tau ,\sigma^r) << 1$. We get a solution of linearized Einstein's
equations, in which the configurational DO $r_{\bar a}(\tau
,\sigma^r)$ play the role of the {\it two polarizations} of the
gravitational wave.

\bigskip

However, the point   canonical transformation of Refs.\cite{7b},
adapted to 13 of the 14 first class constraints is not suited for
the inclusion of matter due to its {\it non-locality}. Therefore the
search started for a local point transformation adapted only to 10
of the 14 constraints, i.e. not adapted to the super-Hamiltonian and
super-momentum constraints.

\bigskip

In Ref.\cite{12b} (inspired by the so-called York - Lichnerowicz
conformal approach \cite{13b} to metric gravity in globally
hyperbolic  space-times based on the decomposition ${}^3g_{ij} =
\phi^4\, {}^3{\hat g}_{ij}$ of the 3-metric with  $\phi = (det\,
{}^3g)^{1/12}$ being the {\it conformal factor}) a new
parametrization of the original 3-metric ${}^3g_{ij}$ was proposed,
which allows to find local point  canonical transformation, adapted
to 10 of the 14 constraints of tetrad gravity, implementing a York
map. The 3-metric ${}^3g_{rs}$ may be diagonalized with an {\it
orthogonal} matrix $V(\theta^r)$, $V^{-1} = V^T$, $det\, V = 1$,
depending on 3 Euler angles $\theta^r$. The gauge Euler angles
$\theta^r$ give a description of the 3-coordinate systems on
$\Sigma_{\tau}$ from a local point of view, because they give the
orientation of the tangents to the 3 coordinate lines through each
point (their conjugate momenta $\pi_i^{(\theta )}$ are determined by
the super-momentum constraints), $\phi$ is the conformal factor of
the 3-metric, i.e. the unknown in the super-hamiltonian constraint
(its conjugate momentum is the gauge variable describing the form of
the simultaneity surfaces $\Sigma_{\tau}$), while the two
independent eigenvalues of the conformal 3-metric ${}^3{\hat
g}_{rs}$ (with determinant equal to 1) describe the genuine {\it
tidal} effects $R_{\bar a}$,  $\bar a = 1,2$, of general relativity
(the non-linear "graviton", with conjugate momenta $\Pi_{\bar a}$).
In the York canonical basis \cite{12b} the gauge variable, which
describes the freedom in the choice of the clock synchronization
convention, i.e. in the definition of the instantaneous 3-spaces
$\Sigma_{\tau}$, is the trace ${}^3K(\tau ,\sigma^r)$ of the
extrinsic curvature of $\Sigma_{\tau}$.

\medskip

The tidal effects $R_{\bar a}(\tau ,\sigma^r)$, $\Pi_{\bar a}(\tau
,\sigma^r)$, are DO {\it only} with respect to the gauge
transformations generated by 10 of the 14 first class constraints.
Let us remark that, if we fix completely the gauge and we go to
Dirac brackets, then the only surviving dynamical variables $R_{\bar
a}$ and $\Pi_{\bar a}$ become two pairs of {\it non canonical} DO
for that gauge: the two pairs of canonical DO have to be found as a
Darboux basis of the copy of the reduced phase space identified by
the gauge and they will be (in general non-local) functionals of the
$R_{\bar a}$, $\Pi_{\bar a}$ variables. This shows the importance of
canonical bases like the York one: the tidal effects are described
by {\it local} functions of the 3-metric and its conjugate
momenta.\medskip

The {\it arbitrary gauge variables} of the York canonical basis are
$\alpha_{(a)}$, $\varphi_{(a)}$, $\theta^i$, $\pi_{\tilde \phi}$,
$n$ and ${\bar n}_{(a)}$. As shown in Refs.\cite{11b,12b}, they
describe the following generalized {\it inertial effects}:

a) the angles $\alpha_{(a)}(\tau ,\sigma^r)$ and the boost
parameters $\varphi_{(a)}(\tau ,\sigma^r)$  describe the
arbitrariness in the choice of a tetrad to be associated to a
time-like observer, whose world-line goes through the point $(\tau
,\sigma^r)$. They fix {\it the unit 4-velocity of the observer and
the conventions for the gyroscopes and their transport along the
world-line of the observer}.

b) the angles $\theta^i(\tau ,\sigma^r)$ (depending only on the
3-metric) describe the arbitrariness in the choice of the
3-coordinates on the simultaneity surfaces $\Sigma_{\tau}$ of the
chosen non-inertial frame  centered on an arbitrary time-like
observer. Their choice induces a pattern of {\it relativistic
standard inertial forces} (centrifugal, Coriolis,...), whose
potentials are contained in the weak ADM energy $E_{ADM}$. These
inertial effects are the relativistic counterpart of the
non-relativistic ones (they are present also in the non-inertial
frames of Minkowski space-time).

c) the {\it shift} functions ${\bar n}_{(a)}(\tau ,\sigma^r )$,
appearing in the Dirac Hamiltonian, describe which points on
different simultaneity surfaces have the same numerical value of the
3-coordinates. They are the inertial potentials describing the
effects of the non-vanishing off-diagonal components ${}^4g_{\tau
r}(\tau ,\sigma^r)$ of the 4-metric, namely they are the {\it
gravito-magnetic potentials} responsible of effects like the
dragging of inertial frames (Lens-Thirring effect) (see the
Ciufolini-Wheeler book in Refs.\cite{13b}) in the post-Newtonian
approximation.

d) $\pi_{\phi}(\tau ,\sigma^r )$, i.e. the York time ${}^3K(\tau
,\sigma^r)$, describes the arbitrariness in the shape of the
simultaneity surfaces $\Sigma_{\tau}$ of the non-inertial frame,
namely the arbitrariness in the choice of the convention for the
synchronization of distant clocks. Since this variable is present in
the Dirac Hamiltonian ($H_D = A_{ADM} + constraints$), it is a {\it
new inertial potential} connected to the problem of the relativistic
freedom in the choice of the {\it instantaneous 3-space}, which has
no non-relativistic analogue (in Galilei space-time there is an
absolute notion of Euclidean 3-space). Its effects are completely
unexplored. For instance, since the sign of the trace of the
extrinsic curvature may change from a region to another one on the
simultaneity surface $\Sigma_{\tau}$, {\it the associated inertial
force in the Hamilton equations may change from attractive to
repulsive in different regions} since $H_D$ knows this sign. These
inertial forces could imply that part of dark matter is an inertial
effect (see Ref.\cite{14b} for a possible gravito-magnetic origin).

e) the {\it lapse} function $N(\tau ,\sigma^r) = 1 + n(\tau ,
\sigma^r )$, the lapse function appearing in the Dirac Hamiltonian,
describes the arbitrariness in the choice of the unit of proper time
in each point of the simultaneity surfaces $\Sigma_{\tau}$, namely
how these surfaces are packed in the 3+1 splitting.

\medskip

The gauge fixing to the extra 6 primary constraints fixes the
tetrads (i.e. the spatial gyroscopes and their transport law). The 4
gauge fixings to the secondary constraints (the super-Hamiltonian
and super-momentum ones) fix ${}^3K$, i.e.the simultaneity
3-surface, and the 3-coordinates on it. The preservation in time of
these 4 gauge fixings generates other 4 gauge fixing constraints
determining the lapse and shift functions consistently with the
shape of the simultaneity 3-surface and with the choice of
3-coordinates on it (here is the main difference with  most of the
approaches to numerical gravity).
\bigskip

To understand better the Hamiltonian distinction between inertial
and tidal effects, i.e. the nature of the general relativistic
effects, a detailed study of the Post-Newtonian solutions of
Einstein's equations adopted by the IAU conventions \cite{15b} for
the barycentric and geocentric celestial reference frames has begun
\cite{16b}. This is no more an academic research, because in a few
years the European Space Agency (ESA) will start the mission ACES
\cite{17b} about the synchronization of a high-precision
laser-cooled atomic clock on the space station with similar clocks
on the Earth surface by means of microwave signals. If the accuracy
of 5 picosec. will be achieved, it will be possible to make a
coordinate-dependent test of effects at the order $1/c^3$, like the
second order Sagnac effect (sensible to Earth rotational
acceleration) and the general relativistic Shapiro time-delay
created by the geoid \cite{18b}. It will be important to find the
Post-Newtonian deviation from Einstein's convention to be able to
synchronize two such clocks and to understand which metrological
protocols have to be used for time dissemination at this level of
accuracy. See also the Einstein Gravity Explorer proposal for a
mission in the framework of ESA Cosmic Vision. The main objective of
these missions will be the high-precision measurement of the
gravitational redshift around the Earth.

\bigskip

Gravitation in the solar system is described in the barycentric
(BCRS) and geocentric (GCRS) non-rotating celestial reference
systems by means of Post-Newtonian solutions of Einstein's equations
for the metric tensor in harmonic gauges codified in the conventions
IAU2000 \cite{15b}. In these conventions it is shown which is the
relativistic structure of Newton (order $1/c^2$) and gravitomagnetic
(order $1/c^3$) potentials to be used either near the Earth or in
the solar system. Near the Earth the relativistic Newton potential
is given mainly by the multipolar expansion of the geopotential
determined by relativistic geodesy.

\medskip

In space missions near the Earth the orbit of the satellite is
evaluated in the GCRS at the level of few cm by using the
relativistic Newton potential  near the Earth and the
Einstein-Infeld-Hoffmann equations at the order $1/c^2$ \cite{4b}.
Instead the trajectories of the clocks in the ground stations in the
GCRS are evaluated from their positions fixed on the Earth crust
(ITRS, International Terrestrial Reference System) by using the
non-relativistic IERS2003 conventions \cite{19b}.

\medskip

For the propagation of rays of light (radar signals) between the
satellite and the Earth stations one uses the null geodesics of the
Post-Newtonian solution in the GCRS.  Till now these geodesics have
been theoretically evaluated \cite{15b} only for the monopole
approximation ($G\, M/ R$) of the geopotential at the order $1/c^2$
in $ct$ (i.e. with only the relativistic Newton potential) and for
an axisymmetric body at the order $1/c^3$ in $ct$ (i.e. with Newton
and gravitomagnetic potentials developed in multipolar expansions)
\cite{20b}. In both cases the potentials are time independent.
Probably further theoretical calculations including the
time-dependence of the potentials will be needed and are under
investigation \cite{16b}.
\medskip

Let $x^{\mu} = \Big(c\, t; \vec x\Big)$ be geocentric 4-coordinates
with $t$ the geocentric time. At the order $1/c^2$ the
post-Newtonian solution in these 4-coordinates is $g_{oo} = 1 -
{{2\, U}\over {c^2}}$, $g_{oi} = 0$, $g_{ij} = - (1 + {{2\, U}\over
{c^2}})\, \delta_{ij}$ (with the convention $(+---)$ for the metric
signature). The instantaneous (non-euclidean at the order $1/c^2$)
3-space is defined by $x^o = c\, t = const.$ The world-line of an
Earth station $B$ is parametrized as $x^{\mu}_B(t) = (x^o_B = c\, t;
{\vec x}_B(t) = r_B(t)\, {\hat x}_B(t) )$ ($r_B = |{\vec x}_B|$),
while the world-line of the satellite $A$ is $x^{\mu}_A(t) = ( x^o_A
= c\, t; {\vec x}_A(t) = r_A(t)\, {\hat x}_A(t) )$. Let ${\vec
v}_C(t) = {{d\, {\vec x}_C(t)}\over {dt}}$ and ${\vec a}_C(t) =
{{d\, {\vec v}_C(t)}\over {dt}}$ be the 3-velocity and
3-acceleration of $C = A,B$.\medskip

If at $t = t_A$ the satellite $A$ emits an electro-magnetic signal,
its reception at the Earth station $B$ will happen at time $t_B >
t_A$ such that $\triangle^2_{AB} = (x_A - x_B)^2 =  [c^2\, (t_A -
t_B)^2 - {\vec N}^2_{AB}] = 0$ with ${\vec N}_{AB}(t_A,t_B) = {\vec
x}_A(t_A) - {\vec x}_B(t_B)\, {\buildrel {def}\over =}\, R_{AB}(t_A,
t_B)\, {\hat N}_{AB}(t_A, t_B)$, ${\hat N}_{AB}^2 = 1$.

Since in real experiments the position ${\vec x}_B(t_A)$ of the
Earth station at the emission time is better known than the position
${\vec x}_B(t_B)$ at the reception time, the quantity $R_{AB}$ has
to be re-expressed in terms of the {\it instantaneous} distance
${\vec D}_{AB} = {\vec x}_A(t_A) - {\vec x}_B(t_A)$, $D_{AB} =
|{\vec D}_{AB}|$. To order $c^{-2}$  in $ct$ (i.e. $c^{-3}$ in $t$)
one gets \cite{15b}

\bea
 R_{AB} &=& | {\vec x}_A(t_A) - {\vec x}_B(t_B)| =
 \sqrt{\Big[{\vec D}_{AB} + {\vec v}_B(t_A)\, R_{AB} + {1\over 2}\,
 {\vec a}_B(t_A)\, R^2_{AB} +O(R^3_{AB}) \Big]^2} =\nonumber \\
 &=& D_{AB} + {1\over c}\, {\vec D}_{AB} \cdot {\vec
 v}_B(t_A) + \nonumber \\
 &+& {1\over {c^2}}\, D_{AB}\, \Big[ {\vec v}_B^2(t_A) +
 {{({\vec D}_{AB} \cdot {\vec v}_B(t_A)}\over {D^2_{AB}}} +
 {\vec D}_{AB} \cdot {\vec a}_B(t_A)\Big]+
  O({1\over {c^3}}).
 \label{1}
 \eea
\medskip

For ACES  the satellite is the Space Station at the altitude of $4\,
10^5\, m$. The relevant quantities are: $v_A = 7.7\, 10^3\, m/s$,
$v_B = v_{ground} = 465\, m/s$, $U({\vec x}_B)/c^2 = 6.9\,
10^{-10}$, $U({\vec x}_A)/c^2 = 6.5\, 10^{-10}$. The Earth
parameters are $G\, M_E = 3.98\, 10^{14}\, m^3/s^2$, $R_E = 6.37\,
10^6\, m$. The experimental uncertainties expected for ACES are at
the level of $5\, ps$ for time transfer and $5\, 10^{-17}$ for
frequency transfer. In the one-way time and frequency transfers one
must add atmosphere-dependent corrections, which tend to be
compensated in the two-way transfers.

\bigskip

The main results of Ref.\cite{15b} are\medskip

A) {\it Gravitational red shift}.\medskip

A1) One-way frequency transfer

\bea
 {{\nu_A}\over {\nu_B}} &=& {{1 - {1\over {c^2}}\, \Big[U({\vec x}_B(t_B)
 + {1\over 2}\, {\vec v}^2_B(t_B)\Big]}\over {1 - {1\over {c^2}}\, \Big[U({\vec x}_A(t_A)
 + {1\over 2}\, {\vec v}_A^2(t_A)\Big]}}\, {{q_A}\over
 {q_B}},\nonumber \\
 &&{}\nonumber \\
 q_A &=& 1 - {{{\vec N}_{AB} \cdot {\vec v}_A(t_A)}\over c} - {{4\, G\, M_E}\over {c^3}}\,
 {{\Big(r_A(t_A) + r_B(t_B)\Big)\, {\vec N}_{AB} \cdot {\vec v}_A(t_A) + R_{AB}\,
 {{{\vec x}_A(t_A) \cdot {\vec v}_A(t_A)}\over {r_A(t_A)}}}\over
 {\Big(r_A(t_A) - r_B(t_B)\Big)^2 - R^2_{AB}}},\nonumber \\
 q_B &=& 1 - {{{\vec N}_{AB} \cdot {\vec v}_B(t_B)}\over c} - {{4\, G\, M_E}\over {c^3}}\,
 {{\Big(r_A(t_A) + r_B(t_B)\Big)\, {\vec N}_{AB} \cdot {\vec v}_B(t_B) + R_{AB}\,
 {{{\vec x}_B(t_B) \cdot {\vec v}_B(t_B)}\over {r_B(t_B)}}}\over
 {\Big(r_A(t_A) - r_B(t_B)\Big)^2 - R^2_{AB}}}.\nonumber \\
 &&{}
 \label{2}
 \eea

For ACES the $J_2$-terms of the geopotential in the factor $q_A/q_B$
do not exceed $4\, 10^{-17}$. One has: 1) first-order Doppler effect
with $|{\vec N}_{AB} \cdot {\vec v}_A/c| \leq 2.6\, 10^{-5}$ for the
satellite and $|{\vec N}_{AB} \cdot {\vec v}_B/c| \leq 1.6\,
10^{-6}$ for the ground; 2) second order Doppler effect with ${\vec
v}^2_A/ 2c^2 \leq 3.4\, 10^{-10}$ for the satellite and ${\vec
v}^2_B/2c^2 \leq 1.3\, 10^{-12}$ for the ground; gravitational
redshift: $U({\vec x}_A)/c^2 = 6.5\, 10^{-10}$ for the satellite and
$U({\vec x}_B)/c^2 = 6.9\, 10^{-10}$ for the ground; 3) the terms of
order $1/c^3$ are less than $3.6\, 10^{-14}$ for the satellite and
$2.2\, 10^{-15}$ for the ground. \medskip

A2) Two-way frequency transfer [$U_{AB} = U({\vec x}_B(t_B)) -
U({\vec x}_A(t_A))$, ${\vec v}_{AB} = {\vec v}_A(t_A) - {\vec
v}_B(t_B)$]

\bea
 {{\triangle\, f}\over f}{|}_{AB} &=& {1\over {c^2}}\, \Big(U_{AB} - {1\over 2}\, {\vec v}^2_{AB} -
 {\vec N}_{AB} \cdot {\vec a}_B(t_B)\Big)\,
 \Big(1 + {{{\hat N}_{AB} \cdot {\vec v}_{AB}}\over c}\Big) +\nonumber \\
 &+& {{R_{AB}}\over {c^3}}\, \Big(- {\vec v}_A(t_A) \cdot {\vec a}_B(t_B) + {\vec N}_{AB} \cdot
 {{d\, {\vec a}_B(t_B)}\over {dt_B}} + 2\, {\vec v}_B(t_B) \cdot {\vec a}_B(t_B) - {\vec v}_B(t_B) \cdot
 \vec \partial\, U_B\Big).\nonumber \\
 &&{}
 \label{3}
 \eea

For ACES one has: 1) gravitational redshift: $U_{AB}/c^2 = 4.6\,
10^{-11}$; second order Doppler effect: $|{\vec v}^2_{AB}/2c^2| \leq
3.3\, 10^{-10}$; 3) acceleration effect: $|{\vec N}_{AB} \cdot {\vec
a}_B/c^2| \leq 7\, 10^{-13}$; 4) pseudo first order Doppler effect:
$|{\hat N}_{AB} \cdot {\vec v}_{AB}/c| \leq 2.7\, 10^{-5}$; 5) last
four terms: they have maximal values $\leq 2\, 10^{-17}, 3.5\,
10^{-19}$ and much less for the last two terms (negligible for
ACES).

\bigskip

B) {\it Shapiro time delay}.\medskip

B1) One-way time transfer

\bea
 T_{AB}(t_A) &=& {1\over c}\, R_{AB} + {{2\, G\, M}\over {c^3}}\, ln\,
 {{|{\vec x}_A(t_A)| + |{\vec x}_B(t_B)| + R_{AB}}\over
 {|{\vec x}_A(t_A)| + |{\vec x}_B(t_B)| - R_{AB}}} =\nonumber \\
 &=&{1\over c}\, D_{AB} + {1\over {c^2}}\, {\vec D}_{AB} \cdot {\vec
 v}_B(t_A) + {1\over {c^3}}\, D_{AB}\, \Big[ {\vec v}_B^2(t_A) +
 {{({\vec D}_{AB} \cdot {\vec v}_B(t_A))^2}\over {D^2_{AB}}} +
 {\vec D}_{AB} \cdot {\vec a}_B(t_A)\Big] +\nonumber \\
 &+& {{2\, G\, M}\over {c^3}}\, ln\,
 {{|{\vec x}_A(t_A)| + |{\vec x}_B(t_A)| + D_{AB}}\over
 {|{\vec x}_A(t_A)| + |{\vec x}_B(t_A)| - D_{AB}}} + O({1\over
 {c^4}})
 \label{4}
 \eea

\medskip

The two terms in $T_{AB}$ beyond $D_{AB}/c$ are usually referred to
as the {\it Sagnac terms} of first ($1/c^2$) and second ($1/c^3$)
order in $t$ due the rotations of the Earth and the satellite. For
ACES (at low elevation of the satellite) they are estimeted to be of
$200\, ns$ for the first order Sagnac term, of $11\, ps$ for the
Shapiro time delay and of $5\, ps$ for the second order Sagnac term.

\bigskip

B2) Two-way time transfer in the T2L2 configuration.

If we consider a signal emitted at $t_A^+$ by the satellite,
reflected at $t_B$ from the Earth station and re-absorbed at $t_A^-$
by the satellite, from the two one-way time transfers $t_B - t_A^+$
and $t_A^- - t_B$ one can get the following estimate of the
deviation from Einstein's $1/2$ clock synchronization convention
\cite{16b} (valid only when one of the two observers A and/or B is
inertial)

\bea
 t_B &=& t_A^+ + {\cal E}\, \Big(t_A^- - t_A^+\Big) = t_A^+ + {\cal E}\, \triangle,\nonumber \\
 &&{}\nonumber \\
 {\cal E} &=& {1\over 2}\, \Big(1 + {1\over c}\, [{\vec v}^+_A + {{\triangle}\over 2}\, {\vec a}^+_A]
 \cdot {\hat n}^+ +\nonumber \\
 &+& {1\over {c^2}}\, \Big[({\vec v}^+_A \cdot {\hat n}^+)^2 - ({\vec v}_A^+)^2 -
 {{G\, M_E\, \triangle}\over {r^+_A}}\, {{({\vec v}_A^+ + {{\triangle}\over 2}\, {\vec a}_A^+)
 \cdot {\hat x}_B(t_b)}\over {{\vec x}_A(t_A^+) \cdot {\hat n}^+ + r_A^+}} \Big]\Big).
 \label{5}
 \eea

\noindent Here ${\hat n}^+$ is the unit tangent to the null geodesic
in the emission point at $t_A^+$, i.e. the direction of the emitted
light (to reach the Earth station one needs consistency with the
orbit determination). ${\hat x}_B(t_B)$ is a unit vector at B. For
ACES, where $t_A^- - t_A^+ \approx 10^{-3}\, s$, the $1/c$ term,
coming from the non-inertiality of both A and B, is the dominating
one: $\approx 10\, ns$. The second order inertial and general
relativistic effects are less than $0.1\, ps$.\medskip

As a consequence there is the possibility to make a  clock
synchronization  taking into account at least the inertial effects
due to the non-inertiality of both Earth stations and satellite.

\bigskip

See Ref.\cite{20b} for the evaluation of the time transfer and the
frequency shift at the order $1/c^4$ in $t$ beyond the monopole
approximation in the case of a stationary post-newtonian
gravitational field generated by an axisymmetric rotating body (the
Earth with constant angular velocity $\vec \omega$), taking also
into account the $\gamma$ and $\beta$ PPN parameters. For Aces, with
time-keeping accuracy $10^{-18}$, there is the evaluation of the
contributions coming from special (Doppler effect) and general (the
first four multipoles of the Earth) to the gravitational redshift.
Instead the gravito-magnetic effects coming from the intrinsic
angular momentum of the Earth are negligible.

\bigskip

Therefore, the problem of light propagation and clock
synchronization is becoming every day more important  due to GPS and
GALILEO \cite{21b}, to the ACES mission \cite{17b}, to the
Bepi-Colombo mission \cite{22b} to Mercury, to the NASA LATOR
proposal \cite{23b}, to GAIA \cite{24b}, to relativistic geodesy
\cite{25b} (determination of the deviation of the geopotential from
the one of the reference ellipsoid from the dependence upon it of
the gravitational redshift) and to the future space navigation
\cite{26b} inside the solar system. The accuracy of the new
generation of atomic clocks \cite{17b} requires a better
understanding of how to perform time dissemination and will induce
deep modifications (like having the standard reference clock in
space) in metrology \cite{27b}.

\end{document}